# An Asynchronous Early Output Full Adder and a Relative-Timed Ripple Carry Adder

P. BALASUBRAMANIAN
School of Computer Engineering
Nanyang Technological University
50 Nanyang Avenue
SINGAPORE 639798
Email: balasubramanian@ntu.edu.sg

*Abstract:* - This article presents the design of a new asynchronous early output full adder which when cascaded leads to a relative-timed ripple carry adder (RCA). The relative-timed RCA requires imposing a very small relative-timing assumption to overcome the problem of gate orphans associated with internal carry propagation. The relative-timing assumption is however independent of the RCA size. The primary benefits of the relative-timed RCA are processing of valid data incurs data-dependent forward latency, while the processing of spacer involves a very fast constant time reverse latency of just 1 full adder delay which represents the ultimate in the design of an asynchronous RCA with the fastest reset. The secondary benefits of the relative-timed RCA are it achieves good optimization of power and area metrics simultaneously. A 32-bit relative-timed RCA constructed using the proposed early output full adder achieves respective reductions in forward latency by 67%, 10% and 3.5% compared to the optimized strong-indication, weak-indication, and early output 32-bit asynchronous RCAs existing in the literature. Based on a similar comparison, the proposed 32-bit relative-timed RCA achieves corresponding reductions in cycle time by 83%, 12.7% and 6.4%. In terms of area, the proposed 32-bit relative-timed RCA occupies 27% less Silicon than its optimized strong-indication counterpart and 17% less Silicon than its optimized weak-indication counterpart, and features increased area occupancy by a meager 1% compared to the optimized early output 32-bit asynchronous RCA. The average power dissipation of all the asynchronous 32-bit RCAs are found to be comparable since they all satisfy the monotonic cover constraint. The simulation results obtained correspond to a 32/28nm CMOS process.

*Key-Words:* - Asynchronous design, Relative-timing, Indication, Ripple Carry Adder, CMOS, Standard cells

## 1 Introduction

The Semiconductor Industry Association's International Technology Roadmap for Semiconductors (ITRS) [1] has identified design for variability as one of the futuristic grand challenges for design technology. The percentage of design reuse in system-on-chip designs which was estimated to be 46% in the 2009 ITRS edition is expected to become 96% by the year 2024. Further, the proportion of design blocks reuse with respect to glue logic is expected to reach 60% by 2024. Moreover, parameter uncertainty as a percentage effect on sign-off delay is expected to increase to 32% by 2024. In this backdrop, a robust flavor of asynchronous design which employs delay-insensitive data codes for data representation and processing, and the 4-phase return-to-zero (RTZ) handshaking protocol for communication is forecast to be a strong contender or an inevitable counterpart to conventional synchronous design for implementing digital circuits and systems in the nanoelectronics regime. This is because robust asynchronous designs[1] are inherently modular (i.e. permits design reuse) [2], are self-checking [3], exhibit superior EMI compatibility [4], are noise-tolerant [5], have the ability to cope with parametric uncertainty, supply voltage, threshold voltage and temperature variations [6] [7], consume power only when and where active [8], and are able to guarantee greater security and robustness against hostile attacks compared to synchronous designs in the case of sensitive industrial applications [9] [10]. Taking cognizance of these facts, the ITRS design report has projected a growing requirement for asynchronous signaling in the nanoelectronics era and also emphasizes on the continuous development of asynchronous circuit and system design tools.

This article presents the design of a new asynchronous early output full adder, which when cascaded leads to an asynchronous RCA that necessitates employing a very small relative-timing assumption to overcome the problem of gate orphan(s) resulting from internal carry propagation. An asynchronous RCA incorporating the relative-timing assumption(s) is called relative-timed RCA.

In the rest of this article, Section 2 provides background information relating to delay-insensitive data encoding, 4-phase RTZ handshaking, and indication (i.e. acknowledgment) in the context of an asynchronous system. The proposed early output asynchronous full adder is presented in Section 3 and its operation is discussed for three possible scenarios viz. carry propagation, carry generation and carry kill. The necessity for imposing relative-timing assumption(s) in an RCA formed using the proposed early output full adder is described in Section 3, and the factual relative-timing assumption required is also estimated. Section 4 presents the simulation results obtained for various 32-bit asynchronous strong-indication, weak-indication, early output, and relative-timed RCAs. The simulation results correspond to power, forward latency, and area. Since the estimation of cycle time of asynchronous logic designs at the gate level is infeasible using a commercial synchronous tool, a theoretical calculation of cycle time of the various 32-bit asynchronous RCAs is presented by assuming different carry propagation lengths within the RCAs and these are also given in Section 4. Finally the conclusions are stated in Section 5.

## 2 Background

An asynchronous logic block comprising an asynchronous digital system represents the combinational logic equivalent of a synchronous digital system [11] [12]. Asynchronous logic blocks constructed using delay-insensitive data codes and adhering to the 4-phase RTZ handshaking protocol are robust.

The dual-rail code is the simplest member of the generic family of delay-insensitive data codes [13], based on which a data wire X is represented using dual data wires X1 and X0 as shown in Fig 1. X = 1 is represented by X1 = 1 and X0 = 0, and X = 0 is represented by X1 = 0 and X0 = 1. These two conditions represent 'valid data', and the condition of both X1 and X0 assuming 0 is referred to as the 'spacer'. The 4-phase RTZ handshaking procedure requires that the application of inputs from the external environment follows the defined sequence: valid data-spacer-valid data-spacer, and so forth.

The representative block diagram of a typical asynchronous system stage is shown in Fig 1 that is accompanied by the sender-receiver analogy. The valid data and spacer processing operations of an asynchronous system stage are explained in [11] [12], and the avid reader is referred to the same. In Fig 1, the junction dots shown enclosed within the blue ovals in dotted lines represent isochronic fork junctions. Isochronicity constitutes the weakest compromise to delay-insensitivity [14], and an isochronic fork junction implies that all the nets forking out from the junction tend to experience similar signal transitions occurring simultaneously.

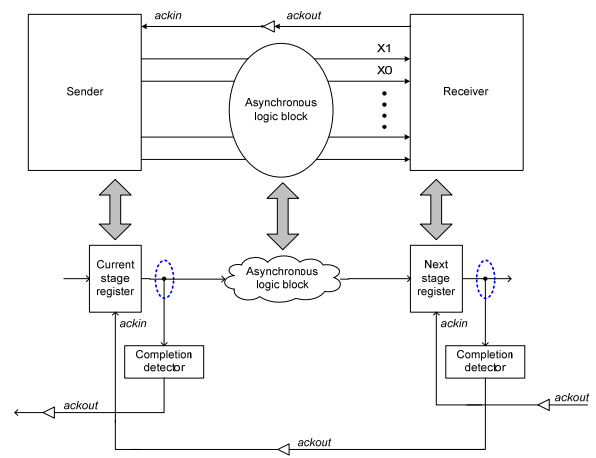

Fig 1. Block diagram of an asynchronous system stage correlated with the sender-receiver analogy

---

[1] Asynchronous design, in this paper, generally refers to a robust flavor which employs delay-insensitive data codes and incorporates 4-phase (return-to-zero) handshaking unless otherwise stated. The term 'robust' may be optionally used as a suffix to emphasize the robustness attribute of the asynchronous design described.

Referring to Fig 1, the 4-phase RTZ protocol is explained as follows. The dual-rail data bus that feeds the current stage register (sender) is initially in the spacer state, and the common acknowledge input (*ackin*) for the current stage register is binary 1, since the common acknowledge output (*ackout*) provided by the next stage register (receiver) is binary 0. The current stage register now transmits a code word, i.e. valid data. This results in low to high transitions on anyone of the corresponding rails of all the dual-rail bus wires which feed the asynchronous logic block. After the next stage register receives a code word subsequent to completion of data processing in the asynchronous logic block, it drives *ackout* to 1, and *ackin* assumes 0. The current stage register waits for *ackin* to become 0 and then resets the data bus, i.e. the data bus feeding the asynchronous logic block is driven to the spacer state. After an unbounded but finite and positive amount of time taken for the resetting of the asynchronous logic block and the passage of spacer to the following register stage, the next stage register drives *ackout* (*ackin*) to 0 (1). With this, a single data transaction is said to be complete, and the asynchronous system is ready to commence the next data transaction.

Asynchronous logic blocks are classified as strongly indicating [15] [16] [49], weakly indicating [15] [17] [49], and early output type [18] [19]. Indication means providing acknowledgment for the receipt of primary inputs through the primary outputs whilst involving the intermediate outputs. With respect to the asynchronous system stage shown in Fig 1, the indication mechanism may be local or global [20]; local – if the asynchronous logic block within the asynchronous system stage by itself indicates the receipt of all the primary inputs, and global – if the asynchronous system stage on the whole indicates the receipt of all the primary inputs by the asynchronous logic block contained within it. It has been recently shown in [21] that local weak-indication is indeed preferable compared to global weak-indication for asynchronous systems from a power-cycle time-area perspective.

The input-output timing relationship of strong-indication, weak-indication, and early output type asynchronous logic blocks is illustrated by the representative timing diagram shown as Fig 2. A strong-indication asynchronous logic block starts to produce the requisite primary outputs only after receiving all the primary inputs whether they are valid data or spacer. A weak-indication asynchronous logic block starts to produce some primary outputs after receiving a subset of the primary inputs. However, the production of at least one primary output is withheld till all the primary inputs are received. The early output asynchronous logic block is in fact more relaxed compared to the strong and weak-indication asynchronous logic blocks in that it could produce all the primary outputs after receiving just a subset of the primary inputs. The early output asynchronous logic block is further subdivided into 2 types as early set and early reset. The early set type asynchronous logic block is capable of producing all the valid primary outputs after receiving just a subset of the valid primary inputs. On the other hand, the early reset type asynchronous logic block is capable of producing all the spacer primary outputs after receiving just a subset of the spacer primary inputs. Hence the early output asynchronous logic block would be categorized as either early set or early reset type, and their respective behaviors are portrayed by the green and pink ovals shown in dotted lines in Fig 2.

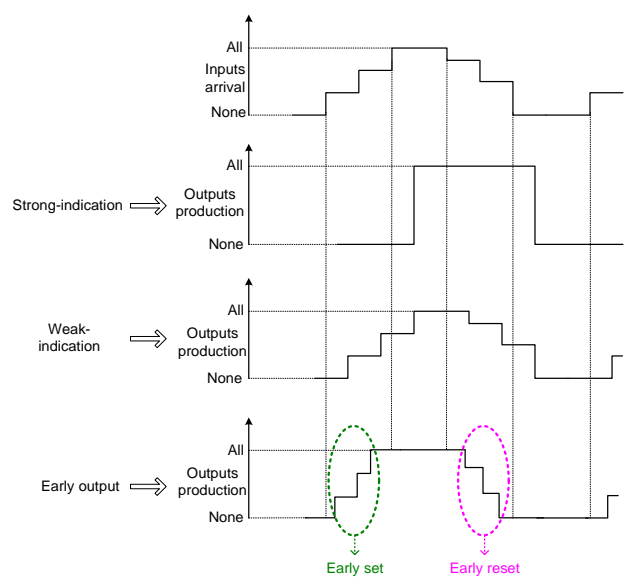

Fig 2. Input-output timing relationship of strong-indication, weak-indication, and early output asynchronous logic blocks

## 3 Proposed Early Output Full Adder and Relative-Timed RCA

The proposed asynchronous early output full adder corresponding to early reset type is shown in Fig 3, which is designed using 11 logic gates. Of these, there are 3 simple gates marked as AND1, AND2 and OR, and 8 complex gates referred by CG1 to CG6 and CE1 and CE2 in Fig 3. Complex gates CG1 to CG5 are referred to as AO22 gates in the digital cell library [22]. The AO22 gate with inputs A, B, C, D and output Y implements the logic function Y = AB + CD. The complex gate CG6 in Fig 3 is referred to as the AO21 gate in the standard cell library [22] and implements the logic function Z = PQ + R, where P, Q, R represent the inputs and Z represents the output. Gates marked CE1 and CE2 represent 2-input C-elements, which are realized using the AO222 gate with feedback [12]. The Muller C-element, introduced in [23], is not only a logic gate but also a state holding register. It outputs 1 only if all its inputs are 1, and likewise outputs 0 only if all its inputs are 0. If its inputs are different, the C-element would maintain its existing state.

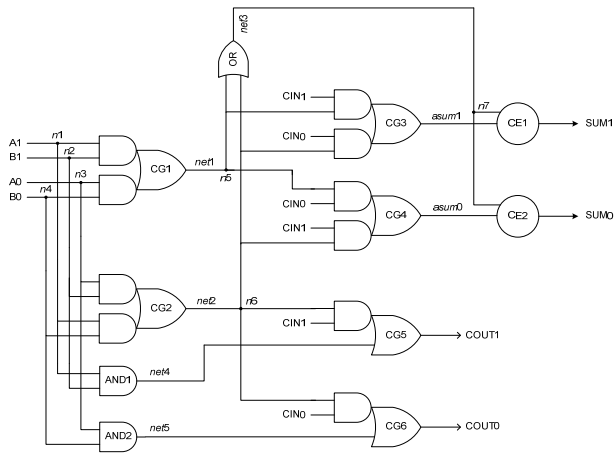

Fig 3. Proposed early output full adder

In Fig 3, (A1, A0), (B1, B0) and (CIN1, CIN0) represent the dual-rail augend, addend and carry inputs, and (SUM1, SUM0) and (COUT1, COUT0) represent the dual-rail sum and carry outputs respectively. The fundamental equations governing the dual-rail sum and carry outputs of the asynchronous full adder are given by (1) to (4).

$$SUM1 = A0B0CIN1 + A0B1CIN0 + A1B0CIN0 + A1B1CIN1 \quad (1)$$

$$SUM0 = A0B0CIN0 + A0B1CIN1 + A1B0CIN1 + A1B1CIN0 \quad (2)$$

$$COUT1 = A0B1CIN1 + A1B0CIN1 + A1B1CIN0 + A1B1CIN1 \quad (3)$$

$$COUT0 = A0B0CIN0 + A0B0CIN1 + A0B1CIN0 + A1B0CIN0 \quad (4)$$

The respective factorized forms [24] of (1) to (4) are specified by (5) to (8).

$$SUM1 = (A0B0 + A1B1) CIN1 + (A0B1 + A1B0) CIN0 \quad (5)$$

$$SUM0 = (A0B0 + A1B1) CIN0 + (A0B1 + A1B0) CIN1 \quad (6)$$

$$COUT1 = (A0B1 + A1B0) CIN1 + A1B1 \quad (7)$$

$$COUT0 = (A0B1 + A1B0) CIN0 + A0B0 \quad (8)$$

Equations (1) to (4) and (5) to (8) are in disjoint sum of products/sum of disjoint products form [25] [26]. A disjoint sum of products expression is formed by the logical disjunction (sum) of disjoint product terms, where any two product terms considered together are mutually disjoint or orthogonal [27], i.e. the logical conjunction of any two products in a disjoint sum of products equation yields null. Hence the degree of mutual orthogonality in relation to two disjoint product terms is at least equal to 1 [35] [49].

The proposed early output full adder shown in Fig 3 synthesizes (5) to (8), and is an example for a circuit with implicit logic redundancy [28]. In Fig 3, nodes $n$1 to $n$7 are isochronic forks, and $net$1, $net$2, $net$3, $net$4 and $net$5 represent the internal outputs. The internal outputs $asum$1 and $asum$0 are logically equivalent to the primary outputs SUM1 and SUM0 respectively. The operation of the proposed early output full adder is described as follows by considering three possible carry scenarios viz. carry propagation, carry generation, and carry kill.

### 3.1 Carry-propagate condition

The carry-propagate condition is specified by either A0 = B1 = 1 or A1 = B0 = 1 during the valid data phase. For either of these input combinations in the valid data phase, the complex gate CG2 will become enabled and the internal output $net$2 will become 1. The OR gate which serves as an internal completion detector will be enabled and $net$3 will also become 1. Assuming CIN1 to be 1, the complex gate CG4 will be activated and the intermediate sum output $asum$0 will also become 1. The C-element, marked as CE2 will be subsequently activated after $asum$0 and $net$3 become 1, leading to the production of 1 on SUM0. With $net$2 and CIN1 becoming 1, the carry output COUT1 will also become 1, i.e. the carry is propagated from input to output. For any of the inputs combination assumed, both the sum and carry outputs are found to be dependent upon the receipt of all the primary inputs. In the successive RTZ phase, even with A0 and/or B1 or A1 and/or B0 assuming 0, both SUM0 and COUT1 will be reset (i.e. returns to 0) regardless of whether CIN1 has returned to 0 or not. This signifies the early output nature of the proposed full adder.

### 3.2 Carry-generate condition

The carry-generate condition is specified by the input combination A1 = B1 = 1 during the valid data phase. For this input combination, AND1 will become enabled and the internal output *net*4 will assume 1. Eventually, the complex gate CG5 will be enabled and COUT1 will be driven to 1 regardless of the receipt of a valid data on the carry input, i.e. the carry is said to be generated from the full adder. Since A1 and B1 are 1, the complex gate CG1 will be activated and *net*1 will assume 1. Subsequently, *net*3 will assume 1. Following this, and depending upon whether CIN0 or CIN1 is 1, SUM0 or SUM1 will assume 1 respectively. Thus for carry generation in the valid data phase, the carry output is not dependent upon the carry input, but the sum output is dependent. In the successive RTZ phase, even with A1 and/or B1 becoming 0, *net*4 will assume 0 and COUT1 will also assume 0. Also, *net*1 will become 0, followed by *net*3 becoming 0, resulting in SUM0 or SUM1 being reset regardless of the carry input returning to 0. This is reflective of the early reset nature of the proposed full adder.

### 3.3 Carry-kill condition

The carry-kill condition is specified by A0 = B0 = 1. For this input combination in the valid data phase, AND2 will be activated and *net*5 will assume 1. This will be followed by an output of 1 on COUT0, i.e. the carry output is said to be killed by the full adder when the input combination A0 = B0 = 1 occurs since the carry is neither generated nor propagated from the input to output. Further, since A0 and B0 are 1, the complex gate CG1 will be activated resulting in *net*1 and *net*3 assuming 1. Given *net*1 = 1, and depending upon whether CIN0 or CIN1 becomes 1, SUM0 or SUM1 will assume 1 respectively. In the subsequent RTZ phase, even if A0 and/or B0 returns to 0, *net*5 will assume 0 and COUT0 will RTZ. Given A0 and/or B0 has returned to 0, *net*1 and *net*3 will assume 0, and eventually SUM0 or SUM1, whichever was 1 earlier will RTZ. Thus in the RTZ phase, both the sum and carry outputs could RTZ regardless of the RTZ of the carry input, which again demonstrates the early reset nature of the proposed full adder.

Having described the early output (early reset) nature of the proposed full adder with respect to carry propagation, carry generation, and carry kill, let us now consider why the need for imposing the relative-timing assumption arises when the proposed full adder is cascaded to form a RCA. We shall consider the 2-bit asynchronous RCA shown in Fig 4 to aid with this discussion.

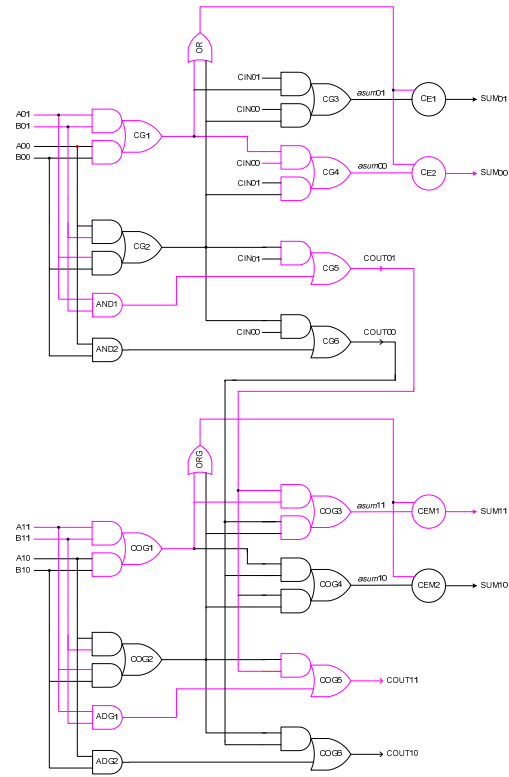

(a) Example application of valid input data in the 2-bit relative-timed RCA
Carry generation: A01 = B01 = 1 with CIN00 = 1 assumed; COUT01 = 1; SUM00 = 1
Carry generation: A11 = B11 = 1; COUT11 = 1; SUM11 = 1 since COUT01 = 1 from the previous stage

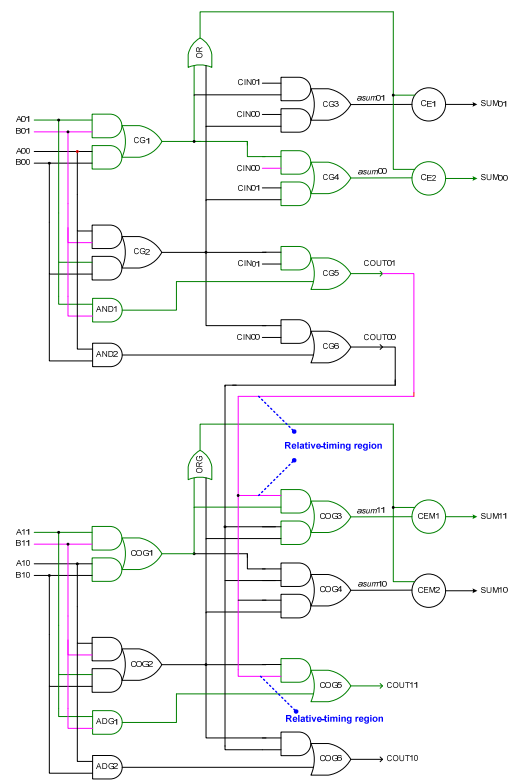

(b) Showing the following RTZ phase in the 2-bit relative-timed RCA with only a partial reset of the primary inputs
RTZ of A01 and A11 alone could facilitate the RTZ of SUM00, COUT01, SUM11 and COUT11
The late RTZ of B01 and B11 would be indicated (acknowledged) by the completion detector
RTZ of COUT01 (also CIN00) may not be acknowledged! Hence, relative-timing assumption for the regions indicated in blue are necessary to avoid the potential problem of gate or wire orphans.

Fig 4. Example application of (a) valid input data and (b) RTZ in the 2-bit relative-timed RCA constructed using the proposed full adder block

In Fig 4, (A01, A00) and (B01, B00) represent the least significant augend and addend inputs, and (A11, A10) and (B11, B10) represent the most significant augend and addend inputs. The carry input for the least significant full adder is represented by (CIN01, CIN00), while the carry input for the most significant full adder is represented by (COUT01, COUT00), which is the carry output of the least significant full adder stage. The least and most significant dual-rail sum outputs are represented by (SUM01, SUM00) and (SUM11, SUM10) respectively. In Fig 4, the complex gates COG1 to COG6 and CEM1 and CME2 synthesize similar logic as that of CG1 to CG6 and CE1 and CE2 respectively. Likewise, the simple gates ADG1, ADG2 and ORG synthesize similar logic as that of AND1, AND2 and OR respectively.

Fig 4a and Fig 4b respectively depict an example application of valid input data and the subsequent RTZ of the 2-bit RCA constructed using the proposed full adder. The pink lines represent valid data (i.e. binary 1) and the green lines represent the RTZ (i.e. binary 0). In Fig 4a, corresponding to the valid input data applied, valid primary outputs are produced. In Fig 4b, as portrayed, even with just a partial RTZ of the primary inputs, all the primary outputs could RTZ. In Fig 4b, the late RTZ of the addend inputs B01 and B11 would not pose a problem since their RTZ would be acknowledged by the completion detector associated with the current stage register as shown in Fig 1. Notice here that the asynchronous logic block of Fig 1 is treated as the 2-bit RCA shown in Fig 4. It is very important to note that during the valid data phase, the production of sum and carry outputs in the RCA would depend upon the actual input pattern applied, while during RTZ, the sum and carry outputs of all the full adder stages within the RCA tend to be reset in parallel due to the early reset nature of the proposed full adder constituting the RCA as shown in Fig 4.

The main issue that now demands attention is resolving the ambiguity associated with ensuring the RTZ of the internal carry signal COUT01 prior to the RTZ of the corresponding sum output viz. SUM11/SUM10. It is essential to presume that COUT01 returns to 0 before SUM11 returns to 0 in Fig 4 to overcome the problem of gate and wire orphans [29] – [31] so as to guarantee the robustness of the RCA. Orphans are basically unacknowledged signal transitions and if present could affect the robustness of an asynchronous circuit/system; for an understanding of gate and wire orphans and their consequences, the interested reader is directed to [29] – [31] for detailed explanations.

For an *n*-bit RCA constructed using the proposed early output full adder, the timing assumption that has to be imposed can be generalized as follows – the carry input of a $(k + 1)^{th}$ stage full adder in the *n*-bit RCA, which is actually the carry output of the $k^{th}$ stage full adder, is assumed to have returned to 0 before the sum output of the $(k + 1)^{th}$ stage full adder in the RCA returns to 0. This signifies a relative-timing assumption [32], and this timing assumption would be independent of the RCA size since the relative-timing assumption is applicable to maximum of only 2 full adder stages within the RCA. This is good news since the relative-timing assumption is restricted to just a small circuit area.

It is important to calculate the actual measure of relative-timing that should be assumed in the RCA with respect to the RTZ phase in order to guarantee the robustness of the relative-timed RCA, a sample of which is shown in Fig 4. To do this, we refer to the cell library information given in [22] and consider only the minimum sized gates in the library since only minimum sized gates were used in the simulations as well.

The maximum propagation delay involved in the direct RTZ of the sum output(s) equals the sum of propagation delays of 3 complex gates viz. 2 AO22 gates and 1 2-input C-element, for example CG1/CG2 and CG3/CG4 and CE1/CE2 in Fig 4 – this equates to a propagation delay of 0.250ns. The maximum propagation delay encountered for the indirect RTZ of the sum output of any full adder stage following the RTZ of the carry output from the previous full adder stage equals 0.313ns. This results from the summation of propagation delays of 2 AO22 gates, 1 AO21 gate and 1 2-input C-element, for example CG2, CG5, COG4 and CEM2 as can be seen in Fig 4. Thus considering the direct RTZ and the indirect RTZ of the sum output of any early output full adder within the relative-timed RCA, it is clear that there is a negative timing slack of 0.063ns, which actually represents the extent of the relative-timing assumption. The worst-case forward latency of the 32-bit relative-timed RCA constructed using the proposed full adder was estimated to be 2.99ns (mentioned in Section 4), and the expected reverse latency is 0.250ns. Compared to these, the required relative-timing assumption of 0.063ns does not appear to be significant. This timing assumption may be reduced further through selective gate sizing [33] of the carry output logic of the full adder shown in Fig 3 and Fig 4.

As mentioned earlier, one of the principal advantages of the relative-timed RCA constructed using the proposed full adder is that the RTZ is very fast and involves a theoretically minimum constant reverse latency of just 1 full adder delay. The forward latency would be data-dependent though. A generalized magnitude-wise comparison of forward and reverse latencies and the cycle time of strong-indication, weak-indication (basic, distributed and biased), early output, and relative-timed $n$-bit RCAs is given in Table 1 [34]. The cycle time is the summation of forward and reverse latencies and determines the throughput of the asynchronous system. In Table 1, $m$ denotes the maximum carry propagation chain length ($m \leq n$), i.e. the maximum number of full adder stages in the RCA through which the carry might propagate.

Table 1. Forward latency, reverse latency, and cycle time attributes of $n$-bit asynchronous RCAs, with $m$ specifying the carry chain length ($m \leq n$)

| RCA type | Forward latency | Reverse latency | Cycle time |
|---|---|---|---|
| Strong | $O(n)$ | $O(n)$ | $O(2n)$ |
| Weak (Basic) | $O(m)$ | $O(m)$ | $O(2m)$ |
| Weak (Distributed/ Biased) | $O(m)$ | $O(2)$ | $O(m+2)$ |
| Early output | $O(m)$ | $O(2)$ | $O(m+2)$ |
| Relative-timed | $O(m)$ | $O(1)$ | $O(m+1)$ |

## 4 Simulation Results and Theoretical Cycle Time Estimates of Various Asynchronous RCAs

Several 32-bit asynchronous RCAs corresponding to strong-indication, weak-indication, early output and relative-timing models were constructed in a semi-custom design fashion using standard cells [22]. The structural integrity of the full adders and the RCAs was preserved during technology mapping which paves the way for a straightforward comparison of their design metrics post-physical synthesis. The 2-input C-element was alone designed manually using the AO222 cell by incorporating feedback and was made available to realize the various asynchronous circuits. High fan-in C-element functionalities wherever likely were safely decomposed into 2-input C-elements based on the safe asynchronous logic decomposition method presented in [35]. Notice that unsafe logic decompositions could result in gate orphans.

An asynchronous system stage comprises the RCA, the input registers, and the completion detector. Of these, the input registers and the completion detector part are identical, and only the RCAs differ in their physical composition. Hence the differences between the simulations results of different asynchronous systems comprising the different RCAs can be entirely attributed to the physical differences in the logical composition of the RCAs.

More than 1000 random input vectors were supplied identically to the RCAs at time intervals of 20ns through test benches in order to capture their switching activities. The value change dump (.vcd) files generated through the functional simulations were used for average power estimation using Synopsys PrimeTime. Since the EDA tool mainly estimates critical path timing, the worst-case forward latency was alone estimated for a typical case PVT specification (1.05V, 25ºC). Appropriate wire loads i.e. parasitic were included whilst performing the simulations. As part of advanced timing analysis, a virtual clock was used just to constrain the input and output ports of the RCAs, and it did not contribute to any power dissipation.

Table 2 presents the simulation results viz. power dissipation, forward latency, and area occupancy of different asynchronous RCAs, each having size of 32-bits. Table 3 lists the recurring logic elements in the critical path of different RCAs.

Table 2. Average power dissipation, forward latency, and area occupancy of different 32-bit asynchronous RCAs estimated using a 32/28nm CMOS process; the corresponding full adder area is given within brackets in the last column

| RCA name and (Indication type) | Power ($\mu$W) | Latency (ns) | Areas ($\mu m^2$) |
|---|---|---|---|
| Reference [36] (Strong) | 2190 | 14.61 | 2529 (54.64) |
| Reference [37] (Strong) | 2181 | 9.26 | 2504.60 (53.88) |
| Reference [38] (Strong) | 2172 | 9.04 | 2293.14 (47.27) |
| Reference [37] (Weak) | 2177 | 8.24 | 2423.27 (51.34) |
| Reference [39] (Weak) | 2192 | 9.66 | 2642.85 (58.20) |
| Reference [40] (Weak) | 2171 | 7.00 | 2016.63 (38.63) |
| Reference [41] (Weak) | 2174 | 4.43 | 2097.96 (41.17) |
| Reference [42] (Weak) | 2171 | 3.32 | 2049.16 (39.65) |
| Reference [19] (Early output) | 2161 | 3.10 | 1658.80 (27.45) |
| This work (Relative-timed) | 2168 | 2.99 | 1675.06 (27.96) |

Table 3. Recurring logic elements in the critical path of different 32-bit asynchronous RCAs corresponding to a 32/28nm CMOS process

| RCA name and (Indication type) | Critical path logic elements |
|---|---|
| Reference [36] (Strong) | 2CE2, 2OR3 |
| Reference [37] (Strong) | CE2, OR4 |
| Reference [38] (Strong) | CE2, 2OR2 |
| Reference [37] (Weak) | CE2, OR3 |
| Reference [39] (Weak) | AND2, CE2, OR3 |
| Reference [40] (Weak) | CE2, OR2 |
| Reference [41] (Weak) | AO222 |
| Reference [42] (Weak) | AO21 |
| Reference [19] (Early output) | AO22 |
| This work (Relative-timed) | AO21 |

ORk – k-input OR gate; ANDk – k-input AND gate; CE2 – 2-input C-element; AO222, AO21 and AO22 are complex gates of the digital standard cell library [22]

From Table 2, it is evident that the average power dissipation of all the asynchronous RCAs is almost comparable – this is because the full adder logic of the RCAs satisfy the monotonic cover constraint [11] [12], which implies the activation of a unique signal path from a primary input to a primary output for each distinct input pattern applied. Among all the RCAs listed in Table 2, the relative-timed RCA incorporating the proposed full adder features the least data path latency, thanks to its optimized data path. Compared to the optimized strongly indicating RCA constructed using the strong-indication full adder of [38], the proposed relative-timed RCA achieves respective reductions in latency and area of 67% and 27%. In comparison with the latency optimized weak-indication RCA constructed using the biased weak-indication full adder of [42], the proposed relative-timed RCA features reduced latency by 10%. In Table 2, Folco et al.'s full adder [40] based RCA features the least area among weak-indication asynchronous RCAs. In comparison with this, the relative-timed RCA incorporating the proposed early output full adder occupies less area by 17%. From Table 2, it can be observed that the relative-timed RCA based on the proposed early output full adder seems to offer a good optimization of power, latency and area simultaneously – thanks to logic factorization and optimization [43] – [47] before physical synthesis.

We shall now discuss the approximate theoretical computation of cycle times of diverse asynchronous RCAs for different carry propagation lengths. A 32-bit asynchronous ALU was implemented in [48] and it was found that addition comprises about 80% of the operations performed by the ALU. It has also been found that about 60% of the random inputs involve carry propagation to less than or equal to 4 stages in the case of 32-bit addition, and almost 100% of the random inputs entail carry propagation to about 8 stages or less. A majority of the address calculations performed by the ALU involve carry propagation of up to 16 stages, and about 45% of the data processing operations require carry-propagation over almost the entire adder width. Given this, an approximate theoretical estimation of the cycle time for different 32-bit asynchronous RCAs by assuming carry propagation lengths of 4, 8, 16, 24 and 28 full adder stages is done as a reflection of the typical operations in an ALU. The calculated cycle times are given in Table 4.

The cycle times are computed by averaging the forward latency values in Table 2 corresponding to the RCA size, and subsequently extrapolating the averaged values based on the generic cycle time magnitudes given in Table 1. As can be seen in Table 4, the relative-timed RCA incorporating the proposed full adder features the least latency for any number of carry propagation stages and on average. In comparison with the optimized strong-indication [38], weak-indication [42], and early output RCAs [19], the proposed relative-timed RCA enables corresponding mean reductions in cycle time by 91%, 16% and 6% respectively.

Table 4. Approximate theoretical cycle time estimates of different 32-bit asynchronous RCAs corresponding to different carry propagation lengths

| RCA type | Cycle time based on number of carry propagation stages | | | | |
|---|---|---|---|---|---|
| | 4 | 8 | 16 | 24 | 28 |
| Reference [36] (Strong) | 29.2 | 29.2 | 29.2 | 29.2 | 29.2 |
| | Mean of cycle times = 29.2ns | | | | |
| Reference [37] (Strong) | 18.5 | 18.5 | 18.5 | 18.5 | 18.5 |
| | Mean of cycle times = 18.5ns | | | | |
| Reference [38] (Strong) | 18.1 | 18.1 | 18.1 | 18.1 | 18.1 |
| | Mean of cycle times = 18.1ns | | | | |
| Reference [37] (Weak) | 2.1 | 4.1 | 8.2 | 12.4 | 14.4 |
| | Mean of cycle times = 8.2ns | | | | |
| Reference [39] (Weak) | 2.4 | 4.8 | 9.7 | 14.5 | 16.9 |
| | Mean of cycle times = 9.7ns | | | | |
| Reference [40] (Weak) | 1.8 | 3.5 | 7.0 | 10.5 | 12.3 |
| | Mean of cycle times = 7.0ns | | | | |
| Reference [41] (Weak) | 0.8 | 1.4 | 2.5 | 3.6 | 4.2 |
| | Mean of cycle times = 2.5ns | | | | |
| Reference [42] (Weak) | 0.6 | 1.0 | 1.9 | 2.7 | 3.1 |
| | Mean of cycle times = 1.9ns | | | | |
| Reference [19] (Early output) | 0.6 | 1.0 | 1.7 | 2.5 | 2.9 |
| | Mean of cycle times = 1.7ns | | | | |
| This work (Relative-timed) | 0.5 | 0.8 | 1.6 | 2.3 | 2.7 |
| | Mean of cycle times = 1.6ns | | | | |

## 5 Conclusions and Future Work

This article has presented a new asynchronous early output full adder design, which when cascaded gives to a RCA that necessitates employing small relative-timing assumptions to guarantee freedom from circuit orphans. The relative-timing assumptions are independent of the RCA size, and are applied on the internal carries to ensure robustness. The primary attributes of the proposed relative-timed RCA are: (i) the time taken to process valid input data is data-dependent, and (ii) the time taken to process the spacer is a constant, and is roughly equal to 1 full adder delay, which is the least possible computation time for spacer.

A 32-bit relative-timed RCA constructed using the proposed early output full adder achieves respective reductions in forward latency by 67%, 10% and 3.5% compared to the optimized strong-indication, weak-indication, and early output 32-bit asynchronous RCAs existing in the literature. Based on a similar comparison, the proposed 32-bit relative-timed RCA achieves corresponding reductions in cycle time by 83%, 12.7% and 6.4%. In terms of area, the proposed 32-bit relative-timed RCA occupies 27% less Silicon than its optimized strong-indication counterpart and 17% less Silicon than its optimized weak-indication counterpart, and features more area occupancy by a meager 1% compared to the optimized early output 32-bit asynchronous RCA. All these simulation results correspond to a 32/28nm bulk CMOS process.

Moreover, the proposed 32-bit relative-timed RCA exhibits the least latency for any number of carry propagation stages and on average. In comparison with the optimized strong-indication, weak-indication, and early output asynchronous RCAs in the literature, the proposed relative-timed RCA enables corresponding mean reductions in cycle time by 91%, 16% and 6% respectively. Thus the relative-timed RCA based on the proposed early output full adder tends to achieve good optimization of power, throughput and area parameters simultaneously.

Scope for future work exists in terms of trying to synthesize efficient early output implementations of single-bit [50] and dual-bit full adder functionality [51] [52] corresponding to both homogeneous and heterogeneous delay-insensitive data encoding, which might lead to better optimized relative-timed RCAs, carry-select adders [53], and multi-operand adders [54], and exploring the design of relative-timed carry lookahead adders [55] [56] [57] by relative-timing the carry lookahead generator in relation to the sum producing logic. All these provide interesting directions for further work.

*References:*
[1] Available: http://www.itrs.net
[2] C.H. van Kees Berkel, M.B. Josephs, S.M. Nowick, "Scanning the technology: applications of asynchronous circuits," *Proceedings of the IEEE*, vo. 87, no. 2, pp. 223-233, February 1999.
[3] I. David, R. Ginosar, M. Yoeli, "Self-timed is self-checking," *Journal of Electronic Testing: Theory and Applications*, vol. 6, no. 2, pp. 219-228, April 1995.
[4] G.F. Bouesse, G. Sicard, A. Baixas, M. Renaudin, "Quasi delay insensitive asynchronous circuits for low EMI," *Proc. 4th International Workshop on Electromagnetic Compatibility of Integrated Circuits*, pp. 27-31, 2004.
[5] N.C. Paver, P. Day, C. Farnsworth, D.L. Jackson, W.A. Lien, J. Liu, "A low-power, low noise, configurable self-timed DSP," *Proc. 4th International Symposium on Advanced Research in Asynchronous Circuits and Systems*, pp. 32-42, 1998.
[6] K.J. Kulikowski, V. Venkataraman, Z. Wang, A. Taubin, M. Karpovsky, "Asynchronous balanced gates tolerant to interconnect variability," *Proc. IEEE International Symposium on Circuits and Systems*, pp. 3190-3193, 2008.
[7] I.J. Chang, S.P. Park, K. Roy, "Exploring asynchronous design techniques for process-tolerant and energy-efficient subthreshold operation," *IEEE Journal of Solid-State Circuits*, vol. 45, no. 2, pp. 401-410, February 2010.
[8] O.C. Akgun, J. Rodrigues, J. Sparsø, "Minimum-energy sub-threshold self-timed circuits: design methodology and a case study," *Proc. 16th IEEE International Symposium on Asynchronous Circuits and Systems*, pp. 41-51, 2010.
[9] Z.C. Yu, S.B. Furber, L.A. Plana, "An investigation into the security of self-timed circuits," *Proc. 9th International Symposium on Asynchronous Circuits and Systems*, pp. 206-215, 2003.
[10] D. Sokolov, J. Murphy, A. Bystrov, A. Yakovlev, "Design and analysis of dual-rail circuits for security applications," *IEEE Transactions on Computers*, vol. 54, no. 4, pp. 449-460, April 2005.
[11] J. Sparsø, S. Furber, *Principles of Asynchronous Circuit Design: A Systems Perspective*, Kluwer Academic Publishers, Boston, MA, USA, 2001.

[12] P Balasubramanian, *Self-Timed Logic and the Design of Self-Timed Adders*, PhD thesis, School of Computer Science, The University of Manchester, 2010.

[13] T. Verhoeff, "Delay-insensitive codes – an overview," *Distributed Computing*, vol. 3, no. 1, pp. 1-8, March 1988.

[14] A.J. Martin, "The limitation to delay-insensitivity in asynchronous circuits," *Proc. 6th MIT Conference on Advanced Research in VLSI*, pp. 263-278, 1990.

[15] C.L. Seitz, "System Timing," in *Introduction to VLSI Systems*, C. Mead and L. Conway (Editors), pp. 218-262, Addison-Wesley, Reading, Massachusetts, USA, 1980.

[16] P. Balasubramanian, D.A. Edwards, "Efficient realization of strongly indicating function blocks," *Proc. IEEE Computer Society Annual Symposium on VLSI*, pp. 429-432, 2008.

[17] P. Balasubramanian, D.A. Edwards, "A new design technique for weakly indicating function blocks," *Proc. 11th IEEE Workshop on Design and Diagnostics of Electronic Circuits and Systems*, pp. 116-121, 2008.

[18] C.F. Brej, J.D. Garside, "Early output logic using anti-tokens," *Proc. 12th International Workshop on Logic and Synthesis*, pp. 302-309, 2003.

[19] P. Balasubramanian, "A robust asynchronous early output full adder," *WSEAS Transactions on Circuits and Systems*, vol. 10, no. 7, pp. 221-230, July 2011.

[20] P. Balasubramanian, N.E. Mastorakis, "Analyzing the impact of local and global indication on a self-timed system," *Proc. 5th European Computing Conference*, pp. 85-91, 2011.

[21] P. Balasubramanian, N.E. Mastorakis, "Global versus local weak-indication self-timed function blocks – a comparative analysis," *Proc. 10th International Conference on Circuits, Systems, Signal and Telecommunications*, pp. 86-97, 2016.

[22] Synopsys Digital Standard Cell Library *SAED_EDK32/28_CORE Databook*, 2012.

[23] D.E. Muller, W.S. Bartky, "A theory of asynchronous circuits," *Proc. International Symposium on the Theory of Switching*, Part I, pp. 204-243, Harvard University Press, 1959.

[24] P. Balasubramanian, R. Arisaka, "A set theory based factoring technique and its use for low power logic design," *International Journal of Computer, Electrical, Automation, Control and Information Engineering*, vol. 1, no. 3, pp. 721-731, 2007.

[25] P. Balasubramanian, R. Arisaka, H.R. Arabnia, "RB_DSOP: a rule based disjoint sum of products synthesis method," *Proc. 12th International Conference on Computer Design*, pp. 39-43, 2012.

[26] P. Balasubramanian, N.E. Mastorakis, "A set theory based method to derive network reliability expressions of complex system topologies," *Proc. Applied Computing Conference*, pp. 108-114, 2010.

[27] P. Balasubramanian, D.A. Edwards, "Self-timed realization of combinational logic," *Proc. 19th International Workshop on Logic and Synthesis*, pp. 55-62, 2010.

[28] P. Balasubramanian, D.A. Edwards, W.B. Toms, "Redundant logic insertion and latency reduction in self-timed adders," *VLSI Design*, vol. 2012, Article ID 575389, pages 13, 2012.

[29] P. Balasubramanian, "Comments on "Dual-rail asynchronous logic multi-level implementation"," *Integration, the VLSI Journal*, vol. 52, no. 1, pp. 34-40, January 2016.

[30] P. Balasubramanian, K. Prasad, N.E. Mastorakis, "Robust asynchronous implementation of Boolean functions on the basis of duality," *Proc. 14th WSEAS International Conference on Circuits*, pp. 37-43, 2010.

[31] C. Jeong, S.M. Nowick, "Block-level relaxation for timing-robust asynchronous circuits based on eager evaluation," *Proc. 14th IEEE International Symposium on Asynchronous Circuits and Systems*, pp. 95-104, 2008.

[32] K.S. Stevens, R. Ginosar, S. Rotem, "Relative timing," *IEEE Transactions on VLSI Systems*, vol. 11, no. 1, pp. 129-140, February 2003.

[33] G. Posser, G. Flach, G. Wilke, R. Reis, "Gate sizing minimizing delay and area," *Proc. IEEE Computer Society Annual Symposium on VLSI*, pp. 315-316, 2011.

[34] P. Balasubramanian, N.E. Mastorakis, "Timing analysis of quasi-delay-insensitive ripple carry adders – a mathematical study," *Proc. 3rd European Conference of Circuits Technology and Devices*, pp. 233-240, 2012.

[35] P. Balasubramanian, N.E. Mastorakis, "QDI decomposed DIMS method featuring homogeneous/heterogeneous data encoding," *Proc. International Conference on Computers, Digital Communications and Computing*, pp. 93-101, 2011.

[36] N.P. Singh, *A Design Methodology for Self-Timed Systems*, M.Sc. thesis, MIT Laboratory

for Computer Science Technical Report TR-258, 1981.

[37] J. Sparsø, J. Staunstrup, "Delay-insensitive multi-ring structures," *Integration, the VLSI Journal*, vol. 15, no. 3, pp. 313-340, October 1993.

[38] W.B. Toms, *Synthesis of Quasi-Delay-Insensitive Datapath Circuits*, PhD thesis, The University of Manchester, 2006.

[39] W.B. Toms, D.A. Edwards, "A complete synthesis method for block-level relaxation in self-timed datapaths," *Proc. 10th International Conference on Application of Concurrency to System Design*, pp. 24-34, 2010.

[40] B. Folco, V. Bregier, L. Fesquet, M. Renaudin, "Technology mapping for area optimized quasi delay insensitive circuits," *Proc. IFIP International Conference on VLSI-SoC*, pp. 146-151, 2005.

[41] P. Balasubramanian, D.A. Edwards, "A delay efficient robust self-timed full adder," *Proc. IEEE 3rd International Design and Test Workshop*, pp. 129-134, 2008.

[42] P. Balasubramanian, "A latency optimized biased implementation style weak-indication self-timed full adder," *Facta Universitatis, Series: Electronics and Energetics*, vol. 28, no. 4, pp. 657-671, December 2015.

[43] P. Balasubramanian, B. Raghavendra, "Analysis of effect of pre-logic factoring on cell based combinatorial logic synthesis," *International Journal of Computer, Electrical, Automation, Control and Information Engineering*, vol. 2, no. 5, pp. 1468-1473, 2008.

[44] P. Balasubramanian, K. Anantha, "Power and delay optimized graph representation for combinational logic circuits," *International Journal of Computer Science*, vol. 2, no. 1, pp. 47-53, 2007.

[45] P. Balasubramanian, R.T. Naayagi, A. Karthik, B. Raghavendra, "Evaluation of logic network representations for Achilles heel Boolean functions," *International Journal of Computers, Systems and Signals*, vol. 9, no. 1, pp. 42-55, 2008.

[46] P. Balasubramanian, R. Chinnadurai, M.R. Lakshmi Narayana, "Minimization of dynamic power consumption in digital CMOS circuits by logic level optimization," *WSEAS Transactions on Circuits and Systems*, vol. 4, no. 4, pp. 257-266, April 2005.

[47] P. Balasubramanian, R. Chinnadurai, M.R. Lakshmi Narayana, "Effecting power consumption reduction in digital CMOS circuits by a hybrid logic synthesis technique," *Proc. 4th WSEAS International Conference on Electronics, Signal Processing and Control*, pp. 132-137, 2005.

[48] J.D. Garside, "A CMOS VLSI implementation of an asynchronous ALU," *Proc. IFIP WG10.5 Working Conference on Asynchronous Design Methodologies*, pp. 181-192, 1993.

[49] P. Balasubramanian, D.A. Edwards, "Power, delay and area efficient self-timed multiplexer and demultiplexer designs," *Proc. IEEE 4th International Conference on Design and Technology of Integrated Systems in Nanoscale Era*, pp. 173-178, 2009.

[50] P. Balasubramanian, D.A. Edwards, C. Brej, "Self-timed full adder designs based on hybrid input encoding," *Proc. 12th IEEE Symposium on Design and Diagnostics of Electronic Circuits and Systems*, pp. 56-61, 2009.

[51] P. Balasubramanian, D.A. Edwards, "Dual-sum single-carry self-timed adder designs," *Proc. IEEE Computer Society Annual Symposium on VLSI*, pp. 121-126, 2009.

[52] P. Balasubramanian, D.A. Edwards, "Heterogeneously encoded dual-bit self-timed adder," *Proc. IEEE PhD Research in Microelectronics and Electronics Conference*, pp. 120-123, 2009.

[53] P. Balasubramanian, C. Jacob Prathap Raj, S. Anandhi, U. Bhavanidevi, N.E. Mastorakis, "Mathematical modeling of timing attributes of self-timed carry select adders," *Proc. 4th European Conference of Circuits Technology and Devices*, pp. 228-243, 2013.

[54] P. Balasubramanian, D.A. Edwards, W.B. Toms, "Self-timed multi-operand addition," *International Journal of Circuits, Systems and Signal Processing*, vol. 6, no. 1, pp. 1-11, 2012.

[55] P. Balasubramanian, D.A. Edwards, H.R. Arabnia, "Robust asynchronous carry lookahead adders," *Proc. 11th International Conference on Computer Design*, pp. 119-124, 2011.

[56] P. Balasubramanian, D.A. Edwards, W.B. Toms, "Self-timed section-carry based carry lookahead adders and the concept of alias logic," *Journal of Circuits, Systems and Computers*, vol. 22, no. 4, pp. 1350028:1-24, April 2013.

[57] P. Balasubramanian, D. Dhivyaa, J.P. Jayakirthika, P. Kaviyarasi, K. Prasad, "Low power self-timed carry lookahead adders," *Proc. 56th IEEE International Midwest Symposium on Circuits and Systems*, pp. 457-460, 2013.